# Temperature, magnetic field, and pressure dependence of the crystal and magnetic structures of the magnetocaloric compound Mn$_{1.1}$Fe$_{0.9}$(P$_{0.8}$Ge$_{0.2}$)

D. M. Liu[1], Q. Z. Huang[2], M. Yue[1], J. W. Lynn[2]*, L. J. Liu[1], Y. Chen[2,3], Z. H. Wu[4], J. X. Zhang[1]

[1]Key Laboratory of Advanced Functional Materials Ministry of Education, Beijing University of Technology, 100 Pingleyuan, Chaoyang District, Beijing 100022, China

[2] NIST Center for Neutron Research, National Institute of Standards and Technology, Gaithersburg, Maryland 20899

[3] Department of Materials Science and Engineering, University of Maryland, College Park, Maryland 20742

[4] Neutron Scattering Laboratory, China Institute of Atomic Energy, P.O. Box 275(30), Beijing 102413, China

ABSTRACT

Neutron powder diffraction studies of the crystal and magnetic structures of the magnetocaloric compound Mn$_{1.1}$Fe$_{0.9}$(P$_{0.8}$Ge$_{0.2}$) have been carried out as a function of temperature, applied magnetic field, and pressure. The data reveal that there is only one transition observed over the entire range of variables explored, which is a combined magnetic and structural transformation between the paramagnetic to ferromagnetic phases ($T_c \approx$ 255 K for this composition). The structural part of the transition is associated with an expansion of the hexagonal unit cell in the direction of the *a*- and *b*-axes and a contraction of the *c*-axis as the FM phase is formed, which originates from an increase in the intra-layer metal-metal bond distance. The application of pressure is found to have an adverse effect on the formation of the FM phase since pressure opposes the expansion of the lattice and hence decreases $T_c$. The application of a magnetic field, on the other hand, has the expected effect of enhancing the FM phase and increasing $T_c$. We find that the substantial range of temperature/field/pressure coexistence of the PM and FM phases observed is due to compositional variations in the sample. *In-situ* high temperature diffraction measurements were carried out to explore this issue, and reveal a coexisting liquid phase at high temperatures that is the origin of this variation. We show that this range of coexisting phases can be substantially reduced by appropriate heat treatment to improve the sample homogeneity.

## II. INTRODUCTION

Magnetic refrigeration based on the magnetocaloric effect (MCE) has attracted recent interest as a potential replacement for the classical vapor compression systems in use today, because it offers potential energy savings, it doesn't use gases harmful to the atmosphere, the system is more compact because the working material is a solid, and its operation should be significantly quieter.[1] In general one wants as large an MCE as possible for the working material, and one way to increase this is to have a system that has a combined magnetic and structural transition, thereby circumventing the limitation imposed by relying only on the magnetic entropy of the system for the thermodynamic cycle. This has led to the *giant* magneto-caloric effect found in $MnAs_{1-x}Sb_x$[2] and $Gd_5Si_2Ge_2$[3] The closely related hexagonal $MnFeP_{1-x}As_x$ compound was found to have a first-order transition from a paramagnetic to a ferromagnetic phase which exhibits a huge MCE.[4] However, the high cost of Gd and the toxicity of As make it unlikely that these compounds will have widespread commercial applications. Recently, it was found that by replacing As with Ge and Si, a large MCE can be achieved near room temperature in a magnetic field ranging from 0 to 5 T [5-12]. In particular, we found[12] that the compound $Mn_{1.1}Fe_{0.9}P_{0.8}Ge_{0.2}$ could be tuned to have a very high MCE, and established that the compound undergoes a combined first-order structural and magnetic phase transition from a paramagnetic (PM) to a ferromagnetic (FM) phase at about 255 K. The two phases were shown to have distinct crystal structures with the same the $Fe_2P$-type hexagonal symmetry, with the magnetic entropy varying directly with the fraction of each phase as a function of magnetic field or temperature. Thus we demonstrated that the MCE was completely governed by the transformation from one crystal and magnetic phase to the other.

In the present study we have used neutron diffraction to analyze the crystal and magnetic structural properties as function of temperature, pressure, and magnetic field, focusing on the region of the PM↔FM transition where the MCE properties are of most interest. The purpose of the analysis is to obtain new results that allow us to elucidate the relationship between the crystal structure, composition, and the magnetic properties in this regime, and understand the nature of the first-order transition in this system. One aspect from the previous work[12] is that the Ge concentration was found to vary macroscopically in the sample, and this varied the transition properties and appeared to inhibit the PM↔FM transformation. Here we identify one important factor that controls the range of coexistence of the two phases, and demonstrate a sample preparation procedure that reduces this range and improves the transformation properties.

## II. EXPERIMENTAL DETAILS

High resolution powder diffraction data were collected at the NCNR on the BT-1 high-resolution neutron powder diffractometer, using monochromatic neutrons of wavelength 1.5403 Å produced by a Cu(311) monochromator. Söller collimations before and after the monochromator and after the sample were 15′, 20′, and 7′ full-width-at-half-maximum (FWHM), respectively. Data were collected in the 2θ range of 3° to 168° with a step size of 0.05° for various temperatures from 300 K to 5 K. Magnetic field measurements were carried out with a vertical field 7 T superconducting magnet, and the structural refinements were performed in the same manner as previously.[12] One sample was the identical one that was used in the earlier study and is the sample used for all the magnetic field, pressure, and most of the temperature-dependent data reported here. Two additional samples of the same nominal composition were prepared for the high temperature annealing studies.

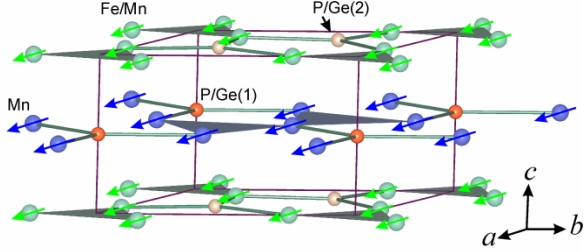

Fig. 1. (color online) Atomic positions and magnetic structure of the ferromagnetic phase. Both the Fe and Mn moments are ordered ferromagnetically, with the moments lying in the *a-b* plane. A magnetic structure with *P11m'* symmetry (moments along the a-axis) was used in the refinements. The refined ordered moments at 10 K are 2.9(1) $\mu_B$ and 1.7(1) $\mu_B$ for the Mn (pyramidal) and the Fe/Mn (tetrahedral) sites, respectively.

Detailed temperature, magnetic field, and pressure-dependent measurements were carried out on the high-intensity BT7 and BT9 triple axis spectrometers. For the pressure measurements, a stainless-steel cell was employed with a maximum pressure of 1.0 GPa, with helium gas as the pressure medium. On both instruments a pyrolytic graphite (PG) (002) monochromator was employed to provide neutrons of wavelength 2.36 Å, and a PG filter was used to suppress higher-order wavelength contaminations. Coarse collimations of 60′, 50′, and 50′ FWHM on BT7 and 40′, 48′, and 40′ FWHM on BT9 were employed to maximize the intensity. No energy analyzer was used in these measurements.

## III. RESULTS and DISCUSSION

$Mn_{1.1}Fe_{0.9}P_{0.8}Ge_{0.2}$ has the $Fe_2P$-type hexagonal structure, space group $P\bar{6}2m$, and the symmetry of the magnetic structure is *P11m'*. As shown in Fig. 1, the Mn atoms are coplanar with the P/Ge(1) atoms, and the Fe/Mn atoms are coplanar with P/Ge(2). The intra-plane transition metals form a triangular configuration. The Mn atoms are surrounded by four P/Ge(2) atoms located on the layers above and below and by one apical P/Ge(1) atom on the same layer, forming a pyramid. The Fe/Mn site is coordinated by two P/Ge(2) atoms located on the same layer and two P/Ge(1) atoms in the layers above and below, forming a tetrahedron.

The data and combined magnetic and structural refinement fit at 10 K, with ambient magnetic field and pressure, are shown in Fig. 2. The agreement between observed and calculated intensities, shown in this figure as an example, is excellent. The quantitative results of the refinements are given in Table 1, where they are compared with the results reported in Ref. [12] for a temperature of 295 K. The relevant bond distances are provided in Table 2.

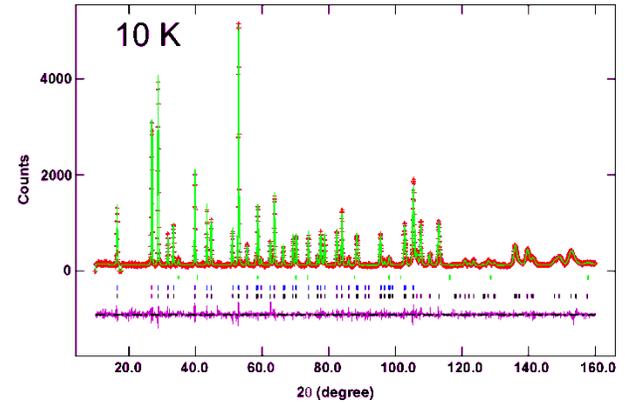

Fig. 2. (color online) Observed (crosses) and calculated (continuous lines) intensities for data collected at 10 K, where only the FM phase is present. Differences are shown in the low part of the plots. Vertical lines indicate the angular positions of the diffraction lines for the nuclear (bottom), magnetic (middle), and impurity (top) structures, respectively. The impurity peaks (due to MnO) were taken into account in the refinements.

The temperature variations of the lattice parameters for the paramagnetic phase (PMP) and ferromagnetic phase (FMP) and the Fe and Mn magnetic moments in the FMP are shown in Fig. 3. We see that the *a*-axis lattice parameter increases and the *c*-axis lattice parameter decreases abruptly at the transition, as previously mentioned. It is readily discerned that, aside from the sharp changes that occur at the phase transition, there is little variation with temperature. The Mn ordered magnetic moments are almost double that of Fe, and both of them increase slightly with the decrease of temperature.

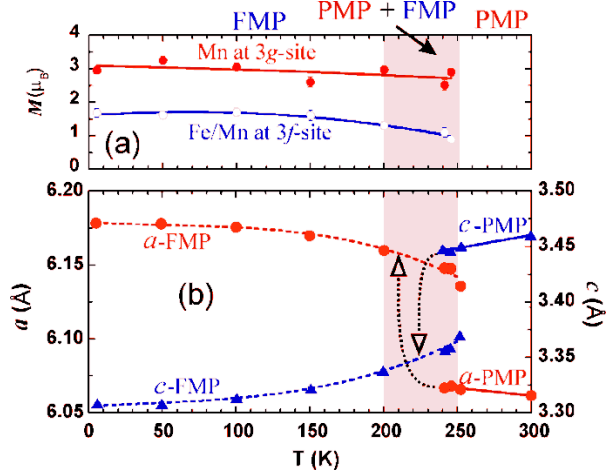

Fig. 3. (color online) Lattice parameters of the PMP and FMP and magnetic moments of Mn and Fe/Mn atoms as a function of temperature. The *c*-axis lattice parameter sharply decreases in going from the PMP to FMP, while the a-axis lattice parameter increases. In the FMP phase the magnetic moment of the Mn does not significantly change over the entire range of temperatures, while the moment on the Fe/Mn site exhibits a very modest decrease at elevated temperatures.

The temperature dependence of the relevant metal-metal bond distances is shown in Fig. 4. It is interesting to note that, while the intra-layer metal-metal bond distances show a significantly large increase in going from the PM to FM phase, the inter-layer distances either remain approximately constant or decrease slightly, in spite of the large decrease of the *c*-axis lattice parameter. The data in Figs. 3 and 4 also show that the shortening of the *c*-axis is mainly due to a decrease of the P/Ge(1)-Mn-P/Ge(1) angle (about 4%).

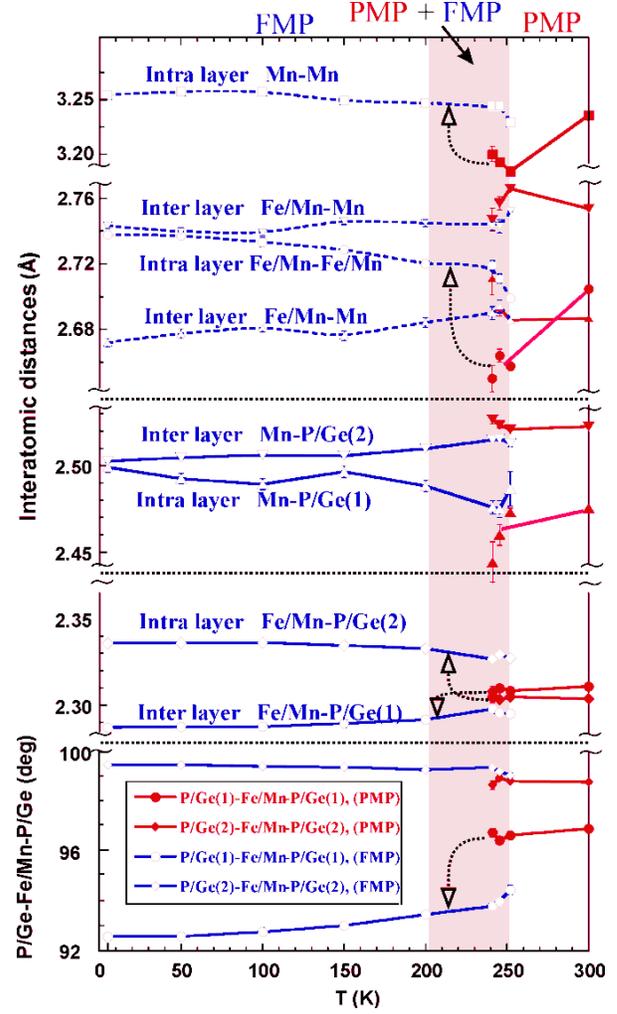

Fig. 4. (color online) Variation of the metal-metal bond distances as function of temperature. The changes occurring at the transition are indicated by the arrows.

Fig. 5 shows the relative atomic positions in the *a-b* plane in the PM and FM phases. The atomic shifts between the two phases are indicated by the arrows, and their effect on the bond distances is clearly visible. In particular, on the z=0 layer, the Fe/Mn-Fe/Mn distances sharply increase, as do the Fe/Mn-P/Ge(2) distances. These atomic readjustments are facilitated by the indicated rotations taking place about the P/Ge(2) atoms. On the z=1/2 layer, the same behavior occurs for the Mn-Mn distances, with only a slight variation of the Mn-P/Ge(1) separations.

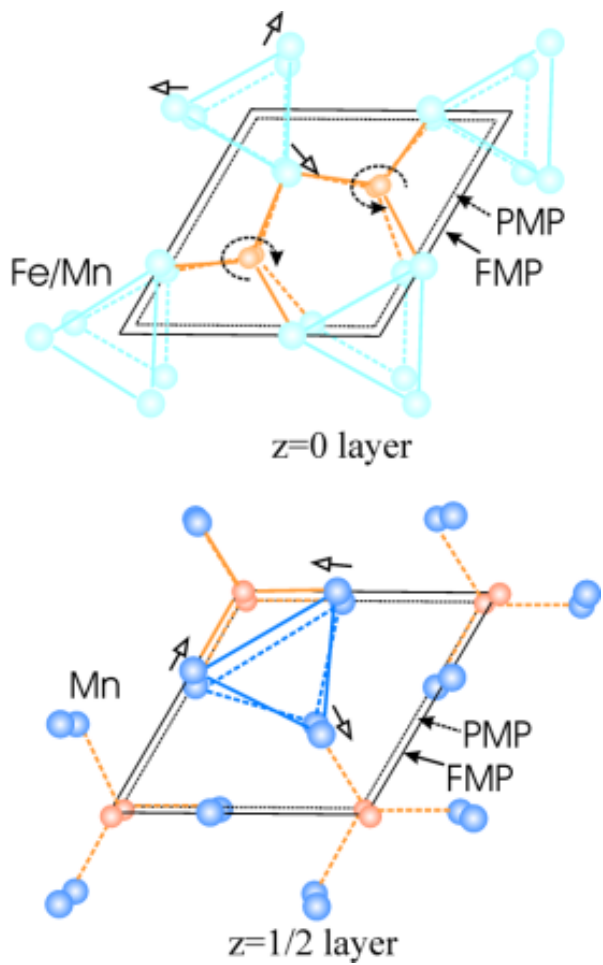

Fig. 5. (color online) Projections along the *c*-axis of the atomic arrangement in the z=0- and z=1/2-layers of the structure. The atomic shifts and the rotations about the P/Ge atoms taking place at the transition are indicated by the arrows. The outlines of the unit cell, and the bonds between the atoms, are shown by continuous and broken lines for the FMP and PMP, respectively.

A close look at the atomic arrangement illustrated in Fig 5 reveals that the bond distances most affected by the phase transformation are the metal-metal distances in both layers, resulting in an expansion of the triangular disposition of the atoms. Since the metal atoms are the ones that are involved in the magnetic coupling, it is clear that the structural and magnetic transitions are intimately related to one another, and therefore any factor that affects these interatomic distances will have an effect, in one direction or the other, on the transition temperature. In particular, from these results we may conclude that the ferromagnetic state is facilitated (and the value of $T_c$ increased) by those factors that tend to cause an expansion in the *a*- and *b*-directions of the cell. This underlines the importance of the type of doping used in an effort to improve the properties of the system. The crystallographic data reveal that changes in the value of $T_c$ can be obtained not only by external factors such as pressure or magnetic field, but also by changes in the internal unit cell provided by altering the composition of the compound. Clearly, the substitution of P with larger atoms like Si or Ge [8] would be advantageous, but there are limitations that have to be considered. For example, replacing P with Ge, which has a much larger covalent radius (1.06 and 1.22 Å, respectively), creates stresses in the structure, and these may have an adverse effect on the value of $T_c$ and on the range of coexistence of the two phases.

The above analysis can also explain the effect of pressure on $T_c$. The application of pressure should inhibit the formation of the FM phase (and decrease $T_c$) because it opposes the expansion of the triangular arrangement of the metal atoms. This is, in fact, what we observe, as shown in Table 3 and in Fig. 6. From Table 3 we find that, at 239 K and ambient pressure, the phase fractions of the PM and FM phases are 19.8 and 80.2%, respectively. At a pressure of 0.92 GPa, however, the phase fraction of the FM phase is reduced to 73.5% and that of the PM phase is increased to 26.5%. Fig. 6 shows the variations of the intensity of the (001) reflections of the PM and FM phases at 245 K, and 0 and 0.92 GPa pressure, as a function of the scattering angle (Fig. 6a) and as function of temperature (Fig. 6b). Note that the transition temperature decreases by

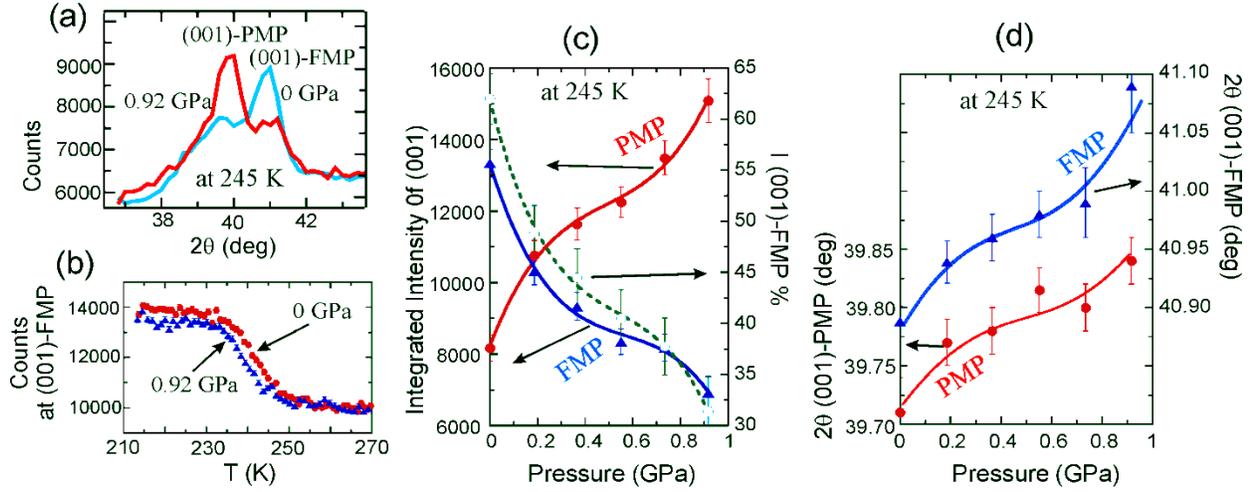

Fig. 6. (color online) Pressure dependence of the scattering measured at 245 K on BT9. (a) (001) peak profiles of the PMP and FMP at 0 (blue line) and 0.92 (red line) GPa. At ambient pressure the intensity of the FM peak (at ≈41°) is higher than that at 0.92 GPa, and the contrary is observed for the intensity of the PM peak. (b) Intensity of the (001) FM peak as a function of temperature for ambient pressure and 0.92 GPa, showing that the transition temperature decreases when pressure is applied. (c) Integrated intensities for the PM and FM reflections as a function of pressure, showing that the intensity of (001)-PM peak increases and that of the (001)-FM peak decreases. The (green) broken line shows the relative intensities of the two peaks. (d) Angular position of the two peaks as a function of pressure. These data show that the lattice parameters of both phases decrease as the pressure increases, as expected.

~3 K under 0.92 GPa. The effect of pressure is particularly evident in Fig. 6c, where the integrated intensity of the PM (001) reflection increases, and that of the FM (001) reflection decreases, as the pressure increases. Moreover, the pressure dependence of the phase transition (Fig. 6c) and (001) peak positions (Fig. 6d) are smooth but nonlinear. The increase in the (001)-PMP intensity and the decrease in the $c$-axis (increasing in 2θ of the (001)-PMP) indicate that the spin ordering is highly correlated to the lattice constants, *i.e.* pressure decreases the metal-metal distances and, therefore, hinders the formation of the FMP.

Fig. 7 shows the effect of an applied magnetic field on the PM-FM transition. The variation of the FMP fraction and the changes of the unit cell volumes of the two phases as a function of the strength of the magnetic field are shown in Fig. 7d, for a temperature of 255 K. As expected, the FMP fraction increases with applied field, which is consistent with the fact that the magnetic field induces the ordering of the spins and therefore favors the formation of the FMP. Note, however, that the transformation to the FM state is not complete even at 7.0 T, with about 20% of the sample remaining in the PMP. We believe this originates from the non-uniformity of the sample that was discovered previously[12], and we will address this issue further below. The lattice parameters also change as the applied field changes (Fig. 7a and 7b), with the cell volume of the PMP decreasing as the field increases while that of the FMP slightly increases. This behavior is consistent with that illustrated in Fig. 3, and will be interpreted in the following discussion. The ordered moments of the Fe and Mn atoms, however, should not change with composition once the transition has taken place, and Fig. 7c shows that this is indeed the case. Comparing Fig. 7c with Fig. 3a, we found that the ordered moments of the Mn ions induced by magnetic field are larger than that caused by temperature, and the ordered moments of the Fe ions show an inverse result. Thus the entropy change caused by a magnetic field is quantitatively different from the entropy change caused by temperature.

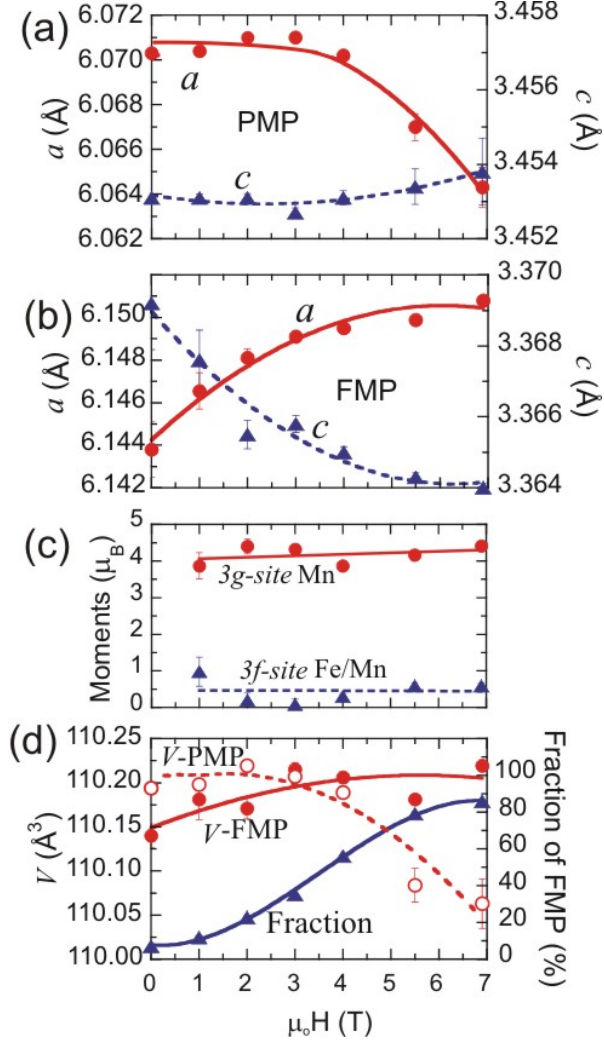

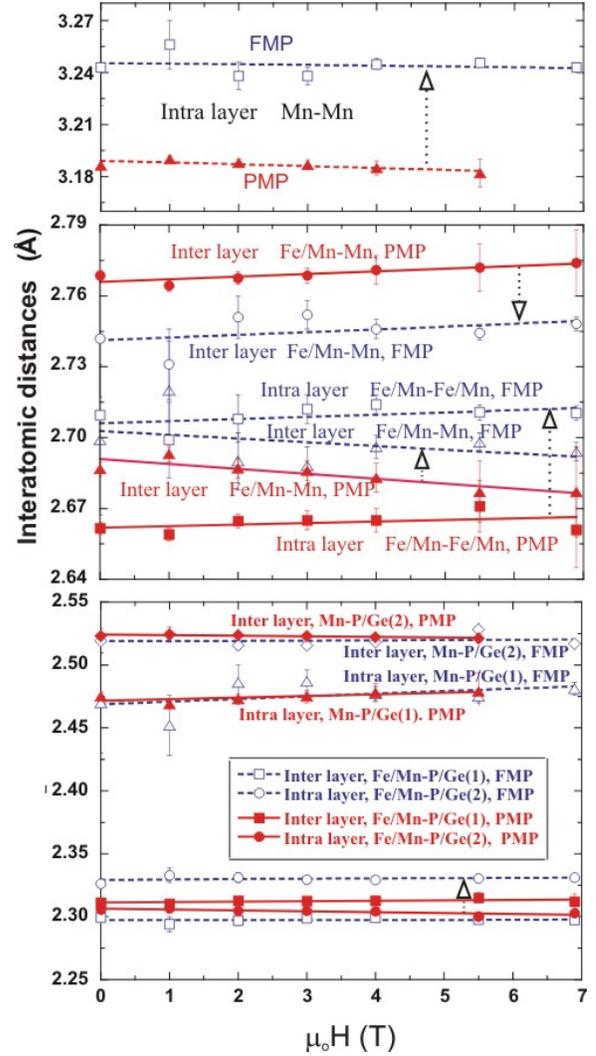

Fig. 7. (color online) Variation of the lattice parameters (Figs 7a and 7b) and of the unit cell volume (Fig. 7d) as a function of applied magnetic field for (a) the PMP and (b) the FMP. (c) The ordered magnetic moments for the Mn and Mn/Fe sites do not vary with field. (d) Fraction of the sample that is ferromagnetic vs. applied magnetic field, and volume of the unit cell for the PMP and FMP.

Fig. 8. (color online) Bond distances at 255 K as a function of applied magnetic field. The intra-layer metal-metal distances show sharp increases in going from the PMP to the FMP, while the inter-layer distances decrease (these variations are indicated by the arrows in the figure). The bond distances of the metal—P/Ge do not show a significant variation with field.

The variation of the bond distances as a function of the applied field is illustrated in Fig. 8. One important aspect to note is that all the distances remain remarkably constant when the field varies, indicating that the crystal and magnetic structures of the PMP and FMP do not change markedly as the transformation progresses, as one would expect in a first-order transition. In going from the PMP to the FMP, the variation of the metal-metal distances is positive and large for the intra-layer bonds, and rather small and negative in the case of the inter-layer bonds. The changes of the intra-layer metal—P/Ge bonds are negligible. This behavior is in general agreement with the behavior shown in Fig. 4, re-enforcing the idea that the effect of temperature on the nature of the transition is basically equivalent to that of an applied magnetic field.

In the PM↔FM transition region, the diffraction results demonstrate unambiguously that the PM and FM phases have different Ge concentrations,[12] which we have attributed as the origin of the wide range of field and temperature over which the two phases are found to coexist. If we assume that different crystallites in the sample have slightly different compositions, they will have also different lattice parameters and convert from one phase to the other at different temperatures, pressures, or magnetic field strengths[13,14]. Consequently, the average lattice parameters will change during the transition as we have observed.

To explore the origin of this compositional inhomogeneity, we have carried out high temperature diffraction measurements on a second of the same nominal composition (designated YPM20). Fig. 9 shows a plot of the observed and calculated intensities collected at 930ºC, using a sample of $Mn_{1.1}Fe_{0.9}P_{0.8}Ge_{0.2}$ fired at that temperature. The pattern shows only the PM structural phase, with no extra lines due to any impurities. This sample was then slowly cooled to 840ºC, and intensity data were collected between 950ºC and 840ºC. The overall scale factors obtained in the refinements at various temperatures are plotted as a function of temperature in Fig.10. The steady increase with decreasing temperature indicates that the volume of the sample is increasing. Thus both liquid and the solid phases are present, with the solid fraction increasing as the cooling takes place. The sample was finally cooled to room temperature, and was found to be in the form of a pellet, also indicating the presence of a (perhaps near-surface) liquid phase. Since Ge (and Mn) enter as solid solution in the $Fe_2P$ structure, we can expect that the compositions of both the solid and liquid phases change during the slow cooling process of the sample, following the lines characteristic of the system's phase diagram. Thus inhomogeneities should be expected unless precautions during preparation are taken.

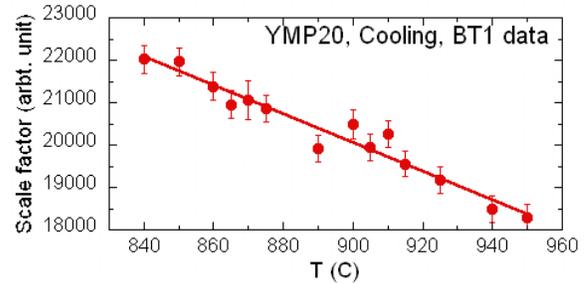

Fig. 10. The overall scale factors obtained in the refinements as a function of temperature.

Based on the results of Fig. 10, a more compositionally uniform sample was prepared by heating the sintered $Mn_{1.1}Fe_{0.9}P_{0.8}Ge_{0.2}$ compound in vacuum quartz tube to 950 ºC quickly, then slowly cooling to 850 ºC over a 24 hour period. The sample was annealed for a further 24 hours at 850ºC, and then quenched in water. Fig. 11 compares the integrated intensities of the (001) neutron reflections for the PM phase of the (first) as-sintered sample (YM01) and the annealed sample (YMA12) on cooling, under identical measurement conditions. The width of the PM↔FM transformation has been reduced almost a factor of two for the annealed sample, from 18 K to 10 K, demonstrating that compositional uniformity is one important factor in optimizing the properties of this system.

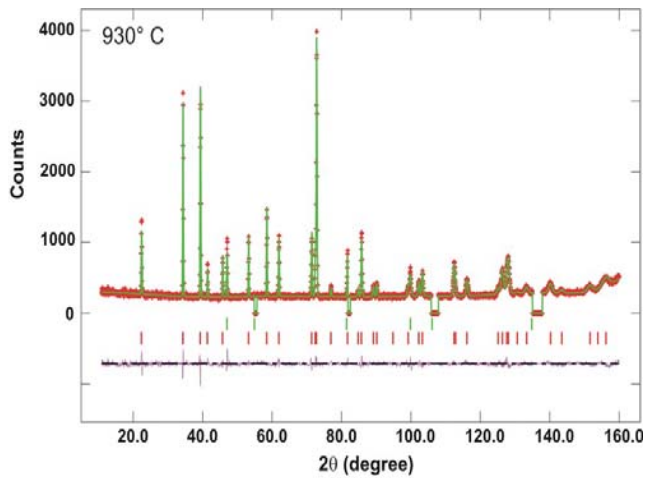

Fig. 9. (color online) Observed (crosses) and calculated (continuous lines) intensities for data collected at 930ºC, where only the PM phase is present. Bragg peaks from the Ta sample holder were excluded from the fit.

We remark that if the PM↔PM transition were continuous (second order) in nature, the magnetic/structural correlation length would grow in the paramagnetic state and diverge at the Curie point. Within the ferromagnetic state, the spin wave excitations would soften and approach zero with a power law behavior as the Curie point is approached. Experimentally we do not see any indication in these diffraction data of the development of either magnetic or structural correlations in the paramagnetic phase that anticipate the combined structural and ferromagnetic transition. On the ferromagnetic side, Fig. 3 shows that there are only small changes in the ordered moments right up to the temperature where the ferromagnetic state collapses, so that temperature produces only a small softening of the ferromagnetism. Both observations indicate that the transition in this system is strongly first order in this giant magnetocaloric system. We note that similar behavior is found in colossal magnetoresistive (CMR) materials such as $La_{1-x}(Ca,Sr,Ba)_xMnO_3$, where the FM↔PM transition is discontinuous [15,16] and is accompanied by structural changes [17,18]. One difference, however, is that the structural changes are on a nanometer length scale rather than long range in nature. The nanoscale structural features make the system especially sensitive to external perturbations such as an applied magnetic field, and it would be interesting to investigate whether the present system possesses such nanoscale correlations. Single crystals likely will be needed to explore this possibility, which might be difficult given the strongly first-order nature of these materials.

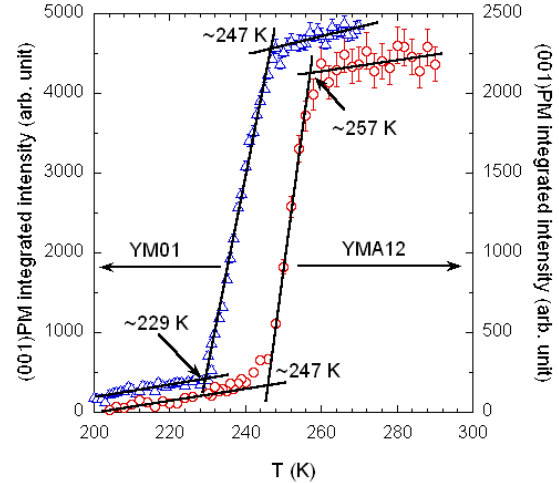

Fig.11 Integrated intensities of the (001) neutron reflections for the PM phase of the original as-sintered sample (YM01) and the newly prepared annealed sample (YMA12) as a function of temperature. Both samples were cooled at the same rate of 15 K/hour.

## IV. SUMMARY AND CONCLUSIONS

The present study emphasizes the importance of the neutron powder diffraction method for understanding the relevant structural details and relating them to the physical properties of both fundamental and technological importance. We have demonstrated that: (i) the phase transition at 255 K is the only one observed in the temperature range from 10 K to 300 K, field range of 0-7 T, and pressure range of 0-1 GPa; (ii) We have shown that Ge doping has the effect of expanding the *a* and *b* lattice parameters of the unit cell and increasing the intra-layer metal-metal bond distances, and this is the origin of the increase of $T_c$; (iii) the application of pressure opposes the expansion of the lattice, and therefore decreases the value of $T_c$; (iv) The application of a magnetic field induces the ordering of the magnetic spins of the Fe and Mn atoms, and thus favors the FM phase and increases $T_c$; (v) Compositional inhomogeneities are responsible for a distribution of $T_c$'s and physical properties. It is demonstrated that appropriate annealing can improve the uniformity of the sample and thereby improve the properties. Further improvement in the MCE properties can be anticipated as fabrication procedures are developed to improve the compositional homogeneity.

## ACKNOWLEDGMENTS

The authors would like to thank Antonio Santoro for many helpful discussions. The work in China was supported by the Key Project of Science &Technology Innovation Engineering, Chinese Ministry of Education (Grant No. 705004).

Table 1. Structural parameters of $Mn_{1.1}Fe_{0.9}P_{0.8}Ge_{0.2}$ at 295 and 10 K. Space group $P\bar{6}2m$. Atomic positions: Mn: $3g(x, 0, 1/2)$; Fe/Mn: $3f(x, 0, 0)$; P/Ge(1): $1b(0, 0, 1/2)$; P/Ge(2): $2c(1/3, 2/3, 0)$.

| Atom | Parameters | 295 K PMP | 10 K FMP |
|---|---|---|---|
| | $a$ (Å) | 6.06137(7) | 6.17811(9) |
| | $c$ (Å) | 3.46023(5) | 3.30669(7) |
| | $V$ (Å$^3$) | 110.098(3) | 109.304(3) |
| Mn | $x$ | 0.5916(3) | 0.5956(5) |
| | $B$(Å$^2$) | 0.77(2) | 0.58(2) |
| | $M$ ($\mu_B$) | | 3.0(1) |
| | $n$(Mn/Fe) | 0.998/0.002(3) | 0.988/0.012(4) |
| Fe/Mn | $x$ | 0.2527(1) | 0.2558(2) |
| | $B$(Å$^2$) | 0.77(2) | 0.58(2) |
| | $M$ ($\mu_B$) | | 1.7(1) |
| | $n$(Fe/Mn) | 0.928/0.072(3) | 0.922/0.078(4) |
| P/Ge(1) | $B$(Å$^2$) | 0.55(4) | 0.54(4) |
| | $n$(P/Ge) | 0.947/0.053(8) | 0.93/0.07(1) |
| P/Ge(2) | $B$(Å$^2$) | 0.55(4) | 0.54(4) |
| | $n$(P/Ge) | 0.726/0.274(4) | 0.736/0.264(6) |
| | $R_P$ (%) | 5.25 | 7.05 |
| | $wR_P$ (%) | 6.65 | 8.75 |
| | $\chi^2$ | 1.276 | 1.913 |

Table 2. Selected interatomic distances (Å) and angles (degree) at 295 K and 10 K.

| | 295 K | 10 K |
|---|---|---|
| *Intra plane metal to metal* | | |
| Mn-Mn | 3.180(1) | 3.254(2) |
| Fe/Mn-Fe/Mn | 2.653(2) | 2.738(2) |
| | | |
| *Inter plane metal to metal* | | |
| Mn- Fe/Mn | 2.686(2) | 2.672(3) |
| Mn- Fe/Mn | 2.771(1) | 2.743(2) |
| | | |
| *Fe/MnP$_4$ tetrahedron* | | |
| Fe/Mn-P/Ge(2) ×2 | 2.3109(6) | 2.3358(7) |
| Fe/Mn-P/Ge(1) ×2 | 2.3039(6) | 2.2874(8) |
| | | |
| *MnP$_5$ pyramid* | | |
| Mn-P/Ge(1) | 2.476(2) | 2.499(3) |
| Mn-P/Ge(2) ×4 | 2.5225(5) | 2.5026(7) |

Table 3. Structural parameters of $Mn_{1.1}Fe_{0.9}P_{0.8}Ge_{0.2}$ at 239 K under an applied pressure of 0.69 GPa. Space group $P\bar{6}2m$. Atomic positions: Mn: $3g(x, 0, 1/2)$; $Fe_{0.928(3)}/Mn_{0.072(12)}$ / $3f(x, 0, 0)$; $P_{0.928(12)}/Ge_{0.072}(1)$: $1b(0, 0, 1/2)$; $P_{0.736(6)}/Ge_{0.264(6)}(2)$: $2c(1/3, 2/3, 0)$.

| Atom | Parameters | 0 GPa | | 0.69 GPa | |
|---|---|---|---|---|---|
| | | PMP 19.8(1)% | FMP 80.2(1) | PMP 26.5(1)% | FMP 73.5(1)% |
| | $a$ (Å) | 6.059(4) | 6.1515(4) | 6.052(1) | 6.1455(4) |
| | $c$ (Å) | 3.47(3) | 3.3555(3) | 3.445(1) | 3.3473(3) |
| | $V$ (Å$^3$) | 109.6(1) | 109.96(2) | 109.30(4) | 109.48(2) |
| Mn | $x$ | 0.64(1) | 0.596(2) | 0.599(6) | 0.603(2) |
| | $B$(Å$^2$) | 0.77(2) | 0.77(2) | 0.6 (1) | 0.6(1) |
| | $M$ ($\mu_B$) | | 4.4(3) | | 4.0(2) |
| Fe/Mn | $x$ | 0.243(5) | 0.2550(8) | 0.253(3) | 0.2538(8) |
| | $B$(Å$^2$) | 0.7(1) | 0.7(1) | 0.6(1) | 0.6(1) |
| | $M$ ($\mu_B$) | | 1.0(2) | | 1.0(2) |
| P/Ge(1) | $B$(Å$^2$) | 0.5(1) | 0.5(1) | 0.6(1) | 0.4(1) |
| P/Ge(2) | $B$(Å$^2$) | 0.5(1) | 0.5(1) | 0.6(1) | 0.4(1) |
| | $R$ (%) | 2.44 | | 2.35 | |
| | $wR$ (%) | 3.13 | | 2.94 | |
| | $\chi^2$ | 2.433 | | 2.129 | |